# To the issue of Geodesics and Torsion in Riemannian geometry and Theory of the Gravitation: analysis of consistence and update of the conventional concept

Yaroslav Derbenev

derbenev@jlab.org

*Jefferson Laboratory, Newport News, Virginia 23606, USA*

*Abstract* A simple differential analysis of issue of the correspondence between notion of geodesics in gravitation theory of GTR and straights of inertial motion in the Minkowski' space-time discovers that, conventional certification of the geodesics in GTR is not compatible with the existence of the Riemann-Christoffel curvature tensor (RCT). We show that, resolution of the contradiction consists of a natural extension of the Christoffel symbols in the gravitation dynamic law to the complete connectedness form which includes a triadic asymmetric tensor (named the *moderator*). The correspondent Riemann supertensor form, unavoidably annihilating by certification of 4-vectors of particle momentum and spin, gives birth to *torsion* (skew-symmetric part of the moderator) and the *gravitensor* (the even-symmetric part); both arrive connected to the RCT and become an indispensable integral part of structure of the gravitational field. The equivalence principle still actual while it becomes enriched in the content. The Einstein-Hilbert law of the metric to matter connection remains unchanged at the produced correction of the gravitational dynamics. Our analysis results in the gravitensor addition to Christoffels in equation for geodesics, modified equations for 4-vector of particle spin with contribution from torsion, and renormalization of metric in the update dynamic concept of the GTR. We pay attention to possible implication of torsion in the elementary interactions.

*Synopsis*

I. Introduction
II. Elements of the covariant differential geometry (§1−2)
III. Euclidian geometry in a curved frame (§ 3)
IV. Riemannian geometry (§ 4 − 6)
V. Geodesics in Riemannian geometry (§ 7 − 8)
**VI.** Gravitation in GTR: review and notes (§ 9)
VII. Update concept of the gravitation field and particle dynamics (§ 10)
VIII. Spin in the update GTR (§ 11)
IX. Resume: Summary and Coclusions

## I. Introduction

Relativistic theory of the gravitation and related dynamics (*general theory of the relativity*, GTR) created by A. Einstein more than a century ago [1, 2] arrived as a 4-fold covariant relativistic generalization of the I. Newton' gravitation law. A start point of the extremely ingenuous investigation undertaken by Einstein was his observation that, effect of the gravitation field is (locally) equivalent of a transition to an accelerated frame (the *equivalence principle*). So not eventually the Riemann−Minkowski geometry with its culture of the covariant differentiation and notion of the *geodesic lines* as an analog of the straight lines of the Euclidian geometry appeared a mathematical basis of the GTR.

There are three different but essentially conjugate aspects of the gravitation theory established by A. Einstein. The first one was the eliciting of nature of the *gravitational field* (GF) as associated with the 4-fold gradient of the space-time *metric tensor*. The second one was establishing of the law of particle motion (form of the *gravitational acceleration*, GA) in the gravitation field, expressed by the rule of the *geodesic lines* (*geodesics*, GLs). The third one was establishing of a law connecting metric tensor to the *matter* (substance with the related inner interactions). This connection in Einstein' concept is expressed in the proportionality of the Riemann-Christoffel tensor (RCT, *curvature form* [3-9]) to the energy-momentum tensor of the matter. Two objects, GF and RCT, in GTR both are structured on the *Christoffel symbols* (CS, or *affine tensor* in terminology by W. Pauli [3]) determined by gradients of the metric.

Transformation law of the CS includes derivatives of the coordinate transformational matrix, therefore it is not a tensor but the *connectedness* object (CO) [3, 8, 9]. It is a mixed valence 3 object with one upper (*contravariant*) index and two down (*covariant*) ones; the CS is symmetric on the latter. As known, the CS transformation law admits a skew-symmetric valence 3 tensor (called *torsion*) as an additive complementary to the CS. However, this addition does not contribute in the GA; probably, for this reason it has not been included in the Einstein' concept and did not play any role in the GTR at all.

By the way, a simple straightforward differential analysis of a background level, presented in this paper, enforces to recognize a question, is a definition of the geodesics traditionally used in GTR actually compatible with existence of the RCT. Namely, analysis in the context of the related comparison with Euclidian (i.e. *flat*) geometry, based on the *correspondence principle*, discovers that it is not. Such an outcome of the undertaken analysis has been puzzling the author of this paper. After a number of the repeated reconsiderations I decided to test an approach to definition of the geodesics with a connectedness object including an unspecified triadic asymmetric tensor (which I call the *tensor-moderator*, TM) as an intrinsic addition to the CSs required for a matching between the geodesic law and RCT. To my satisfaction in principle, I found this approach working with a complete consistence, resulting in establishing of the TM connection to the RCT. The analysis and the results followed by a conclusive summary are constituting the content of the present paper.

## II. Elements of the covariant differential geometry

### § 1. Covariant differentiation

Covariant differentiation of vector and tensor functions [3-9] is an elementary background establishment of the differential geometry; it is inquired in physics at formulation of dynamic laws in form of connection of derivatives of vector and tensor objects to themselves or other ones. Condition of the covariance requires preserving of this form with respect of arbitrary transformations of the space-time coordinates.

*1.1. Vector and tensor functions*

Vector and tensor functions as *geometrical objects* are certified by form of their transformations at transformation of coordinates $x^m$ with differential matrix $A_m^{m'}$:

$$x^m \to x^{m'}(x); \quad k,l,m,n,p,q = 1,2,3,4; \tag{1.1}$$

$$A_m^{m'} \equiv \frac{\partial x^{m'}}{\partial x^m} \equiv A; \quad det\ A \neq 0; \quad \frac{\partial x^m}{\partial x^{m'}} \equiv A_{m'}^m \equiv A^{-1}; \tag{1.2}$$

$$A_n^{m'} A_{n'}^n = \Delta_{n'}^{m'} = \begin{cases} 0; & m' \neq n' \\ 1; & m' = n' \end{cases}; \quad \to \quad A_n^{m'} \partial_k A_{n'}^n = -A_{n'}^n \partial_k A_n^{m'}; \quad \partial_k \equiv \frac{\partial}{\partial x^k}. \tag{1.3}$$

As usual, here and overall below we imply summation on the coinciding upper and down indices.

$$X^{m'} = A_m^{m'} X^m; \qquad X_{m'} = A_{m'}^m X_m; \tag{1.4}$$

$$X_{n'}^{m'} = A_m^{m'} A_{n'}^{n} X_n^m ; \tag{1.5}$$

$$X^{m'n'} = A_m^{m'} A_n^{n'} X^{mn}; \tag{1.6}$$

$$X_{m'n'} = A_{m'}^{m} A_{n'}^{n} X_{mn}, \tag{1.7}$$

$$X_{n'k'}^{m'} = A_{k'}^{k} A_{n'}^{n} A_m^{m'} X_{nk}^m . \quad etc. \tag{1.8}$$

Here $X^m$, $X_m$ and $X_n^m$ are notations for *contravariant* and *covariant* vector and mix valence 2 tensor, respectively; $X^{mn}$ and $X_{mn}$ are the valence 2 the *contra-* and *co-variant* tensors, etc.

*1.2. Connectedness issue of the differential laws*

A differential law connects derivatives of vectors and tensors to these objects themselves or other ones. At transformation with constant matrices $A_m^{m'}$ (*linear transformations*) the derivatives arrive tensors. However, non-linear transformations (generally inquired in differential geometry and GTR) cause appearance of derivatives of $A_m^{m'}$ at the objects. To compensate for breaking of the covariance by these "uninvited" terms, one has to employ *covariant derivatives* (CDs).

*1.2.1. Covariant derivatives and the connectedness objects*

Covariant differentiation is achieved by adding a mix valence 3 *connectedness* object (CO) $G_{nk}^m$ (our notation), generally asymmetric on the down indices [8, 9]), to differential symbol $\partial_k$ :

$$\partial_k X^m \rightarrow \nabla_k X^m \equiv \partial_k X^m + G_{nk}^m X^n; \tag{1.9}$$

$$\partial_k X_m \rightarrow \nabla_k X_m \equiv \partial_k X_m - G_{mk}^n X_n; \tag{1.10}$$

$$\partial_k X_n^m \rightarrow \nabla_k X_n^m \equiv \partial_k X_n^m + G_{pk}^m X_n^p - G_{nk}^p X_p^m , \tag{1.11}$$

$$\partial_k X^{mn} \rightarrow \nabla_k X^{mn} \equiv \partial_k x^{mn} + G_{pk}^m X^{pn} + G_{pk}^n X^{mp}; \tag{1.12}$$

$$\partial_k X_{mn} \rightarrow \nabla_k X_{mn} \equiv \partial_k X_{mn} - G_{mk}^p X_{pn} - G_{nk}^p X_{mp}; \tag{1.13}$$

$$\partial_k X_{nl}^m \rightarrow \nabla_k X_{nl}^m \equiv \partial_k X_{nl}^m - G_{lk}^p X_{np}^m + [\mathbf{G}_k, \mathbf{X}_l], \quad etc.; \tag{1.14}$$

in the last formula, brackets [,] denote commutator of two matrices on indices $m, n$, for which we introduce bold notations:

$$\mathbf{G}_k = G_{nk}^m; \quad \mathbf{X}_l = X_{nl}^m. \tag{1.15}$$

When considering covariant derivatives (CD) as tensors, one can derive the well-known transformation law of the connectedness at transformation of coordinates, which can be written in the following view :

$$G_{nk}^m \rightarrow G_{n'k'}^{m'} = A_{k'}^{k} A_m^{m'} (G_{nk}^m + \Delta_n^m \partial_k) A_{n'}^{n} , \tag{1.16}$$

or, in two equivalent forms:

$$A_{m'}^{m} A_n^{n'} A_k^{k'} G_{n'k'}^{m'} = G_{nk}^m + A_n^{n'} A_k^{k'} \partial_{k'} A_{n'}^{m} , \tag{1.17}$$

or:

$$A_n^{n'} A_k^{k'} G_{n'k'}^{m'} = A_m^{m'} G_{nk}^m - \partial_k A_n^{m'} \equiv \frac{\partial x^{m'}}{\partial x^m} G_{nk}^m - \frac{\partial^2 x^{m'}}{\partial x^n \partial x^k}. \qquad (1.18)$$

Note that, once a particular representation for $G_{nk}^m$ satisfying transformation law (1.16) has been certified, all possible other ones will distinct from it in a triadic tensor $T_{nk}^m$, generally asymmetric on down indexes; this follows directly from transformation formula (1.16). In other words, difference of two distinguish CO is tensor.

*1.2.2. Torsion.* Further, it also follows from equation (1.16) that, the skew-symmetric part of $G_{nk}^m$:

$$\frac{1}{2}(G_{nk}^m - G_{kn}^m) \equiv \overline{T}_{nk}^m = -\overline{T}_{kn}^m \qquad (1.19)$$

transforms as tensor (called *torsion* [3-9], for which we use notation $\overline{T}_{nk}^m$), so one may search for a particular utilization of the connectedness as a form even-symmetric on the down indices; then object $G_{nk}^m$ can be generally represented as sum of the found the even-symmetric one (generally denoted $\overline{\overline{G}}_{nk}^m$) and torsion:

$$G_{nk}^m = \overline{\overline{G}}_{nk}^m + \overline{T}_{nk}^m; \qquad \overline{\overline{G}}_{nk}^m = \overline{\overline{G}}_{kn}^m, \qquad (1.20)$$

with same transformation law for $\overline{\overline{G}}_{nk}^m$ as generally for $G_{nk}^m$:

$$A_n^{n'} A_k^{k'} \overline{\overline{G}}_{n'k'}^{m'} = \frac{\partial x^{m'}}{\partial x^m} \overline{\overline{G}}_{nk}^m - \frac{\partial^2 x^{m'}}{\partial x^n \partial x^k}. \qquad (1.21)$$

*1.3. The Riemannian, or metrical spaces*

*1.3.1. General certification of a Riemannian space*

According to definition accepted in the differential geometry, *Riemannian, or metrical space* is a manifold of free variables (coordinates) $x^k$ with introduced a non-degenerate symmetric tensor $w_{lm}(x)$ (*metric tensor*) [8]:

$$w_{lm} = w_{ml}; \qquad det w_{lm} \neq 0. \qquad (1.22)$$

*1.3.2. Christoffel symbols.* There is a well-known specific solution for symmetric connectedness employed in GTR, the *Christoffel symbols* (CSs) $\Gamma_{nk}^m$, determined by a condition that, covariant derivative of metric tensor $w_{lm}$ with connectedness $\Gamma_{nk}^m$ is equal zero [3-9]:

$$\partial_k w_{lm} - \Gamma_{lk}^n w_{nm} - \Gamma_{mk}^n w_{nl} \equiv \hat{\partial}_k w_{lm} = 0. \qquad (1.23)$$

Condition (2.23), as known, allows one to express the CSs via derivatives of metric tensor:

$$\Gamma_{nk}^m = \frac{1}{2} w^{ml}(\partial_n w_{kl} + \partial_k w_{nl} - \partial_l w_{kn}) = \Gamma_{kn}^m \equiv \mathbf{\Gamma}_k; \qquad (1.24)$$

here $w^{ml}$ is tensor inverse to $w_{ml}$:

$$w^{ml} w_{nl} = \Delta_n^m. \qquad (1.25)$$

The CSs thus appear an analytical functional of metric tensor, a form built on derivatives of the $w_{mn}$.

*1.3.3. General utilization of the connectedness*

The CO (1.20) can be finally represented as follows:

$$G_{nk}^m \Rightarrow \widehat{G}_{nk}^m \equiv \Gamma_{nk}^m + T_{nk}^m; \qquad T_{nk}^m = \bar{T}_{nk}^m + \bar{\bar{T}}_{nk}^m; \qquad \bar{\bar{T}}_{nk}^m = \bar{\bar{T}}_{kn}^m, \qquad (1.26)$$

where $\bar{\bar{T}}_{nk}^m$ is the even-symetric part of tensor $T_{nk}^m$. We also will use notation in bold symbols as matrices on indices $m, n$:

$$G_{nk}^m \equiv \mathbf{G}_k \Rightarrow \widehat{\mathbf{G}}_k \equiv \mathbf{\Gamma}_k + \mathbf{T}_k; \qquad \mathbf{T}_k = \bar{\mathbf{T}}_k + \bar{\bar{\mathbf{T}}}_k, \qquad (1.27)$$

where $\mathbf{T}_k$ is tensor $T_{nk}^m$ generally asymmetric on the down indices. We call combined form (1.26) the *moderate* or *complete* connectedness (MC), and triadic tensor $T_{nk}^m$ the *tensor-moderator* or simply *moderator* (MT). The CSs (1.24) can be characterized as *primal* connectedness, and associated covariant derivatives with $\mathbf{G}_k \Rightarrow \mathbf{\Gamma}_k$ can be called the *primal* CDs (PCDs). CDs with the *moderate* CO (3.27) can be called the *complete* CDs (CCD). In present paper we disclose the MT genesis and connection to the Christoffel symbols.

*1.3.4. Operators of the covariant differentiation (CDO)*

1. General CDO symbol, $\nabla_k$:

$$\nabla_k = \partial_k (+) \mathbf{G}_k; \qquad \mathbf{G}_k \equiv G_{nk}^m. \qquad (1.28)$$

2. Primal CDO, $\hat{\partial}_k$:

$$\hat{\partial}_k \equiv \partial_k (+) \mathbf{\Gamma}_k; \qquad \mathbf{\Gamma}_k \equiv \Gamma_{nk}^m. \qquad (1.29)$$

3. Moderate CDO

$$\widehat{\nabla}_k = \partial_k (+) \widehat{\mathbf{G}}_k = \hat{\partial}_k (+) \mathbf{T}_k; \qquad \widehat{\mathbf{G}}_k = \mathbf{\Gamma}_k + \mathbf{T}_k; \qquad \mathbf{T}_k \equiv T_{nk}^m \qquad (1.30)$$

### § 2. Riemann-Christoffel tensor (RCT), and certification of the metrical geometries

*2.1. RCT*

*2.1.1. RCT form.* In geometry of Riemannian (i.e. *metrical*) spaces a key role belongs the Riemann- Christoffel *curvature form,* tensor (RCT):

$$R_{n;kl}^m \equiv \mathbf{R}_{kl} \equiv \partial_k \mathbf{\Gamma}_l - \partial_l \mathbf{\Gamma}_k + [\mathbf{\Gamma}_k, \mathbf{\Gamma}_l] = -\mathbf{R}_{lk}. \qquad (2.1)$$

Here symbol [,] means commutator of $\Gamma_{nk}^m$ and $\Gamma_{nl}^m$ considered as matrixes on indexes $m, n$.

*2.1.2. Exposition of the RCT.* It is known that, RCT form can be exposed via consideration of the alternate second covariant derivatives of vector functions. Namely, let us assume the existence of a contra- and co-variant vector functions, $V^m$ and $U_m$, respectively. Since their PCD forms:

$$\hat{\partial}_k V^m \equiv \partial_k V^m + \Gamma_{nk}^m V^n \qquad (2.2)$$

$$\hat{\partial}_k U_m \equiv \partial_k U_m - \Gamma_{mk}^n U_n, \qquad (2.3)$$

are tensors, one can derive the following relations, using general formulas (1.11) and (1.13) [8,9]:

$$\hat{\partial}_k \hat{\partial}_l V^m - \hat{\partial}_l \hat{\partial}_k V^m = R_{n;kl}^m V^n; \qquad (2.4)$$

$$\hat{\partial}_k \hat{\partial}_l U_m - \hat{\partial}_l \hat{\partial}_k U_m = -R_{m;kl}^n U_n. \qquad (2.5)$$

The RCT arrives as tensor form, once $\mathbf{\Gamma}_k$ has been established as a connectedness form transformed according to law (1.16), and vice versa.

*2.1.3. Condition for exposing the RCT.* Here, we have to note that, represented exposition of RCT as a non-zero form should be forestalled by a condition that, PCD of vector functions should not be equal zero:

$$\hat{\partial}_l V^m \neq 0; \quad \hat{\partial}_l U_m \neq 0, \tag{2.6}$$

Otherwise, i.e. at:

$$\hat{\partial}_l V^m = 0, \quad and\ (or) \quad \hat{\partial}_l U_m = 0, \tag{2.7}$$

one, apparently, has to acknowledge that, $\mathbf{R}_{kl}$ should be equal zero in such a case, as well. Below we will present an analysis in detail of this issue.

*2.2. Certification of the metrical spaces*

*2.2.1. Euclidian space (ES).* This space is characterized (defined) by a condition that,

$$\mathbf{R}_{kl} \equiv \partial_k \mathbf{\Gamma}_l - \partial_l \mathbf{\Gamma}_k + [\mathbf{\Gamma}_k, \mathbf{\Gamma}_l] = 0. \tag{2.8}$$

Among the coordinate frames of this space there are, obviously, such ones that:

$$\Gamma^n_{mk} = 0 \tag{2.9}$$

over the space; hence, metric tensor $w_{lm}$ is constant (denote it $E_{l_0 m_0}$) in space; we call such frames the Cartesian ones. Curved coordinate frames can be obtained by any non-degenerate transformations with space-dependent matrices:

$$A^l_{l_0} = \frac{\partial x^l}{\partial x^{l_0}}; \quad A^{l_0}_l = \frac{\partial x^{l_0}}{\partial x^l}. \tag{2.10}$$

Then:

$$E_{l_0 m_0} \rightarrow \bar{w}_{lm}(x) \equiv A^{l_0}_l(x) A^{m_0}_m(x) E_{l_0 m_0}; \tag{2.11}$$

here $A^{l_0}_l$ is matrix inverse to matrix $A^l_{l_0}$ of transformation from *Cartesian frame* $x_0$ to frame $x$. Transformation (2.10) generates CSs (1.24) according to law (1.16):

$$\Gamma^m_{nk} \Rightarrow \bar{\Gamma}^m_{nk} \equiv -A^{k_0}_k A^{n_0}_n \partial_{k_0} A^m_{n_0} \equiv -A^{k_0}_k A^{n_0}_n \frac{\partial^2 x^m}{\partial x^{k_0} \partial x^{n_0}}. \tag{2.12}$$

while tensor form (2.1) remains zero, hereby representing equations (2.8) as relations between CSs and their derivatives.

*2.2.2. The strict Riemannian space (SRS).* The SRS and related *Riemannian geometry* is certified with existence of a non-zero *Riemann-Christoffel tensor form* (2.1):

$$\mathbf{R}_{kl} \equiv \partial_k \mathbf{\Gamma}_l - \partial_l \mathbf{\Gamma}_k + [\mathbf{\Gamma}_k, \mathbf{\Gamma}_l] \neq 0. \tag{2.13}$$

Existence of a non-zero RCT form (2.31) is sign of the SRS; once $\mathbf{R}_{kl} \neq 0$, either at a point or over the space, there does not exist a transformation which could turn $\mathbf{R}_{kl}$ to zero. Metric form of an SRS, $w_{mn}(x)$, and CSs (1.24) then cannot be reduced to view (2.11) and (2.12), respectively, as well.

## III. Euclidian geometry in a curved frame

### § 3. Differential geometry of Euclidian spaces

*3.1. Differential law of straights in Euclidian geometry*

*3.1.1. Straights law in Cartesian frame of ES.* A background element of the Euclidian geometry in terms of a Cartesian frame are the *straight* lines $x^m(\tau)$, where $\tau$ is a common *canonical parameter* (note for a clearness that, parameter $\tau$ is not included in collection of free coordinates but is introduced for convenience and universality of representation of the *world lines*). In terms of the differential method, the straights are characterized by their *directions* i.e. *tangent vectors* $g^{m_0}$ constant in space:

$$\partial_{k_0} g^{m_0} = 0; \tag{3.1}$$

$$x^{m_0}(\tau) = g^{m_0}\tau + const. \tag{3.2}$$

*3.1.2. Space dispersion of vectors in a curved frame of ES.* Constant vectors of straights become functions of coordinates in a curved frame:

$$g^{m_0} = const \quad \rightarrow \quad g^m(x) = A^m_{m_0}(x) g^{m_0}. \tag{3.3}$$

Dispersive gradients of vectors in terms of the curved frame can be expressed via the related CSs (2.12):

$$\partial_k g^m(x) = g^{m_0} \partial_k A^m_{m_0} \equiv A^{m_0}_n A^{k_0}_k (\partial_{k_0} A^m_{m_0}) g^n = A^{n_0}_n A^{k_0}_k \frac{\partial^2 x^m}{\partial x^{k_0} \partial x^{n_0}} g^n \equiv -\bar{\Gamma}^m_{nk} g^n \tag{3.4}$$

*3.1.3. Equations for vectors of straights in ES in terms of a curved frame.* Formula (3.4) of ES can be written as equation for covariant derivative of vector $g^m(x)$:

$$\partial_k g^m + \bar{\Gamma}^m_{nk} g^n = 0. \tag{3.5}$$

Equation (3.5) is a covariant identity of the elementary definition of straights in Euclidian space (3.1); we call equation (3.5) the *elementary covariant equation of ES*.

*3.1.4. Elementary equations for straight lines in ES in a curved frame*

First, we remind that, in terms of a curved frame of ES, vector $g^m$ of a straight, generally, acquires gradients over coordinates $x^k$, i.e. becomes function of $x^k$:

$$g^m \Rightarrow g^m(x) = A^m_{m_0}(x) g^{m_0}. \tag{3.6}$$

Considering vector function $g^m(x)$ as tangent vector of a world line $x^m(\tau)$, we write:

$$\frac{dx^m}{d\tau} = g^m(x). \tag{3.7}$$

Next, we take derivative of this equation on parameter $\tau$:

$$\frac{d^2 x^m}{d\tau^2} = \frac{d}{d\tau} g^m[x(\tau)] = \frac{\partial g^m}{\partial x^k} \frac{dx^k}{d\tau} \equiv g^k \partial_k g^m. \tag{3.8}$$

Finally, substituting derivative $\partial_k g^m$ from equations (3.5), we obtain differential law for straight lines of a ES in terms of a curved coordinate frame:

$$\frac{d^2x^m}{d\tau^2} = -\bar{\Gamma}^m_{nk}(x)g^n g^k \equiv -\bar{\Gamma}^m_{nk}(x)\frac{dx^n}{d\tau}\frac{dx^k}{d\tau}. \tag{3.9}$$

*3.2. Primal covariant equation (PCE)*

*3.2.1. Issue of the PCE.* Above in paragraph *2.1.* we pointed that, exposition of RCT (2.1) from process (2.4), (2.5) should imply existence of vector functions $V^m(x)$ and $U_m(x)$, *primal* CD of which (PCD) is *not equal zero*:

$$\hat{\partial}_k V^m \equiv \partial_k V^m + \Gamma^m_{nk}V^n \neq 0; \tag{3.10}$$

$$\hat{\partial}_k U_m \equiv \partial_k U_m - \Gamma^n_{mk}U_n \neq 0. \tag{3.11}$$

Below we will show that, in fact, existence of vector functions subordinate of the laws:

$$\hat{\partial}_k V^m \equiv \partial_k V^m + \Gamma^m_{nk}V^n = 0, \quad \text{and (or)} \quad \hat{\partial}_k U_m \equiv \partial_k U_m - \Gamma^n_{mk}U_n = 0, \tag{3.12}$$

with $\Gamma^m_{nk}$ given by general formula (2.24), is not compatible with existence of the non-zero RCT; actually, this vector object encloses and represents differential geometry of Euclidian space where RCT is zero, together with solution for metric tensor given by formula (2.11).

We call equations (3.12) the *primal covariant equations* (PCE).

*3.2.2. <u>Statement 1</u>.* Equations (3.12) enclose in themselves certificate of the Euclidian space: $\boldsymbol{R}_{kl} = 0$.

*Proof.* Let us consider the alternate second ordinary derivative of vector function $V^m(x)$, a subject of equation (3.12). Since:

$$\partial_k \partial_l V^m - \partial_l \partial_k V^m \equiv 0,$$

it follows immediately from vector equation (3.12), that:

$$\partial_k(\Gamma^m_{nl}V^n) - \partial_l(\Gamma^m_{nk}V^n) = 0; \tag{3.13}$$

so we receive the following relations:

$$(\partial_k \Gamma^m_{nl})V^n - (\partial_l \Gamma^m_{nk})V^n + \Gamma^m_{nl}\partial_k V^n - \Gamma^m_{nk}\partial_l V^n = 0. \tag{3.14}$$

Referring, again, to equation (3.12), we replace derivatives of $V^n$ in (3.14) according to (3.12), and obtain the following relations:

$$R^m_{n;kl}V^n = 0, \tag{3.15}$$

where we introduced notation $R^m_{n;kl}$ for a form structured on the CSs (1.24):

$$R^m_{n;kl} \equiv \partial_k \Gamma^m_{nl} - \partial_l \Gamma^m_{nk} + \Gamma^m_{pk}\Gamma^p_{nl} - \Gamma^m_{pl}\Gamma^p_{nk}, \tag{3.16}$$

which is nothing else but the Riemann-Christoffel tensor form (2.1).

A correspondent procedure applied to equation for covariant vector function $V_m$ in (3.12):

$$\hat{\partial}_k V_m \equiv \partial_k V_m - \Gamma^n_{mk}V_n = 0, \tag{3.17}$$

leads to a similar result:

$$R^n_{m;kl}V_n = 0. \tag{3.18}$$

Equations (3.15) and (3.18) can be referred to arbitrary direction of the autonomic vector objects $V^m$ and $V_m$ at arbitrary point, and overall components of form (3.18), therefore we have to conclude:

$$R^n_{m;kl} = 0, \qquad (3.19)$$

or, in bold symbols:

$$R^m_{n;kl} \equiv \boldsymbol{R}_{kl} = \partial_k \boldsymbol{\Gamma}_l - \partial_l \boldsymbol{\Gamma}_k + [\boldsymbol{\Gamma}_k, \boldsymbol{\Gamma}_l] = 0. \qquad (3.20)$$

We thus have received a direct and immediate outcome from the presented revision of the primal form of the presupposed vector laws (3.12): they are incompatible with existence of the RCT, − i.e. incompatible with the Riemannian geometry.

*3.2.3. Statement 2 .* Primal covariant equation (3.12 )with general CSs form (1.24) appear a complete identity of the rudimentary covariant equation (3.5) of ES.
This statement is a direct consequence of the fact (3.20); proof consists of chain of the logic exposed above in paragraph *2.4.1*.

*3.2.4. Equations for straight lines of ES in a curved frame.* These equations can be derived directly from the primal equations (3.12). Obviously, they arrive identical to equations (3.9) with replacement $\bar{\Gamma}^m_{nk} \implies \Gamma^m_{nk}$:

$$\frac{d^2 x^m}{d\tau^2} = -\Gamma^m_{nk} \frac{dx^n}{d\tau} \frac{dx^k}{d\tau}; \qquad (3.21)$$

however, due to property (3.20), metric $w_{lm}$ appears the literal identity to form (2.11); then, here: $\Gamma^m_{nk} = \bar{\Gamma}^m_{nk}$.

*3.2.5. Conclusion to paragraph 3.2.* Basing on the exposed analysis, one may state the following.

1. Riemannian geometry excludes existence of vector functions with zero PCD (3.12).
2. Primal covariant equations (PCE) (3.12) with CSs (1.24) enclose in themselves certification of Euclidian space (2.8) together with general solution (2.11) for metric form relevant to the Euclidian geometry.
3. PCEs together with equations (3.21) for the correspondent world lines appear equations for straights in Euclidian space in terms of the curved frames.

## IV. Riemannian geometry

### § 4. Vector-mediators of a strict Riemannian space

As it has been pointed above, exposition of RCT (2.1) from process (2.4), (2.5) implies existence of vector functions $V^m$ and (or) $U_m$, *primal* CD of which (PCD) is *not equal zero*:

$$\hat{\partial}_k V^m \equiv \partial_k V^m + \Gamma^m_{nk} V^n \neq 0, \qquad (4.1)$$

and (or) :

$$\hat{\partial}_k U_m \equiv \partial_k U_m - \Gamma^n_{mk} U_n \neq 0, \qquad (4.2)$$

otherwise, second covariant derivatives appear turned to zero, and vector functions cannot serve the exposition of the RCT form. In the context of such a service, we call vectors $V^m$ and $U_m$ the *mediators*. Since the PCDs of both are tensors, we have to presume that:

$$\hat{\partial}_k V^m = \mathrm{X}^m_k; \qquad \hat{\partial}_k U_m = \mathrm{X}_{mk}, \qquad (4.3)$$

where $X_k^m$ and $X_{mk}$ are some tensors. In an important for us disposition of RG as a mathematical base of the GTR, mediators should belong a group of basic objects having the autonomy status; then tensors $X_k^m$ and $X_{mk}$ should be composed as conversions of our vectors with a triadic tensor $T_{nk}^m$, generally asymmetric:

$$X_k^m \Rightarrow -X_{nk}^m V^n; \qquad X_{mk} \Rightarrow X_{mk}^n U_n; \qquad (4.4)$$

then:

$$\hat{\partial}_k V^m = -X_{nk}^m V^n; \qquad \hat{\partial}_k U_m = X_{mk}^n U_n, \qquad (4.5)$$

or:

$$\partial_k V^m = -G_{nk}^m V^n; \qquad \partial_k U_m = G_{mk}^n U_n; \qquad G_{mk}^n \equiv \Gamma_{mk}^n + X_{nk}^m. \qquad (4.6)$$

Triadic tensor $X_{nk}^m$ thus can be identified with the above certified tensor-moderator $T_{nk}^m$ of the *complete connectedness* $\widehat{G}_{nk}^m$ (1.26):

$$X_{nk}^m \Rightarrow T_{nk}^m; \quad \rightarrow \quad G_{nk}^m \Rightarrow \widehat{G}_{nk}^m; \qquad (4.7)$$

then we acquire equations for dual pair of vector functions $V^m(x)$ and $U_m(x)$:

$$\partial_k V^m = -\widehat{G}_{nk}^m V^n; \qquad \partial_k U_m = \widehat{G}_{mk}^n U_n; \qquad \widehat{G}_{nk}^m \equiv \Gamma_{nk}^m + T_{nk}^m, \qquad (4.8)$$

or:

$$\partial_k V^m + \Gamma_{nk}^m V^n = -T_{nk}^m V^n; \qquad \partial_k U_m - \Gamma_{mk}^n U_n = T_{mk}^n U_n. \qquad (4.8)$$

Note that, choice of a relevant structure of tensors $Y_k^m$ and $Z_{mk}$ can be regard as minimal, providing *dual correspondence* of equations for vector functions $V^m$ and $U_m$, with property of constancy of *norm* of $V^m$ and $U_m$ over space following straightforward from moderate equations (4.8):

$$\partial_k (V^m U_m) \equiv 0; \quad \rightarrow \quad V^m U_m = const. \qquad (4.9)$$

Also underline that, the mediators have been certified as the autonomic vector functions, direction (orientation) of which has not been fixed by some external factors. Existence of vector functions $V^m(x)$, $U_m(x)$, subordinate of the exposition law (4.9), appears an intrinsic attribute of the Riemannian geometry.

Presumed equations (4.8) respond a necessity of existence of vector functions with a non-zero PCD indicated in paragraph *2.1.3.* . One may preclude now that, primal equations (3.12) enclose in themselves Euclidian geometry while not entering the strict Riemannian space.

We call statement of the RCT incompatibility with existence of zero PCD vector functions the *mediator theorem 1*.

## § 5. Eliciting of the tensor-moderator

*5.1. Riemann supertensor form (RSF)*

*5.1.1. General certification of the Riemann tensor.* As known, applying procedure of the alternate second covariant derivatives with a generally certified operator $\nabla_k$ designated in equations (1.9) through (1.14), to some vector functions $\breve{V}^m$ and $\breve{U}_m$, one obtains:

$$(\nabla_k \nabla_l - \nabla_l \nabla_k)\breve{V}^m = G^m_{n;kl}\breve{V}^n + (G^n_{kl} - G^n_{lk})\nabla_n \breve{V}^m ; \tag{5.1}$$

$$(\nabla_k \nabla_l - \nabla_l \nabla_k)\breve{U}_m = -G^n_{m;kl}\breve{U}_n + (G^n_{kl} - G^n_{lk})\nabla_n \breve{U}_m ; \tag{5.2}$$

here:

$$\nabla_k \breve{V}^m \equiv \partial_k \breve{V}^m + G^m_{nk}\breve{V}^n; \qquad \nabla_k \breve{U}_m \equiv \partial_k \breve{U}_m - G^n_{mk}\breve{U}_n; \tag{5.3}$$

$$G^m_{n;kl} \equiv \mathbf{G}_{kl} \equiv \partial_k \mathbf{G}_l - \partial_l \mathbf{G}_k + [\mathbf{G}_k, \mathbf{G}_l] = -\mathbf{G}_{lk}; \qquad \mathbf{G}_k \equiv G^m_{nk}; \tag{5.4}$$

*5.1.2. Certification of the Riemann supertensor form.* Since $G^m_{nk}$ is supposed to transform according to law (1.16), it can be considered identity of the naturalized form of the complete connectedness. Structure of form $\mathbf{G}_{kl}$ as functional of the connectedness $G^m_{nk}$ maintains with extension of the affine tensor $\mathbf{\Gamma}_k$ to the complete connectedness form $\widehat{\mathbf{G}}_k$:

$$\mathbf{G}_k \Rightarrow \widehat{\mathbf{G}}_k = \mathbf{\Gamma}_k + \mathbf{T}_k ; \qquad \mathbf{G}_{kl} \Rightarrow \widehat{\mathbf{G}}_{kl} \equiv \widehat{G}^m_{n;kl} \equiv \partial_k \widehat{\mathbf{G}}_l - \partial_l \widehat{\mathbf{G}}_k + [\widehat{\mathbf{G}}_k, \widehat{\mathbf{G}}_l] ; \tag{5.5}$$

$$(\widehat{\nabla}_k \widehat{\nabla}_l - \widehat{\nabla}_l \widehat{\nabla}_k)\breve{V}^m = \widehat{G}^m_{n;kl}\breve{V}^n + 2\overline{T}^n_{kl}\widehat{\nabla}_n \breve{V}^m ; \tag{5.6}$$

$$(\widehat{\nabla}_k \widehat{\nabla}_l - \widehat{\nabla}_l \widehat{\nabla}_k)\breve{U}_m = -\widehat{G}^n_{m;kl}\breve{U}_n + 2\overline{T}^n_{kl}\widehat{\nabla}_n \breve{U}_m ; \tag{5.7}$$

In the context of the general mathematical characterization of tensor form $\widehat{\mathbf{G}}_{kl}$, we call it the *Riemann supertensor form* (RSF). It can be regard as a curl-covariant derivative of the complete connectedness $\widehat{\mathbf{G}}_k$. The RST arrives as tensor form, once $\widehat{\mathbf{G}}_k$ has been established as a connectedness form transformed according to law (1.16), and vice versa. The complete CO $\widehat{\mathbf{G}}_k$ can be regard as the *exhaustive* representation of the connectedness object.

*5.1.3. Conditioning of exposition of the RSF.* We have to note that, similar to the case of the RCT exposition from the above shown procedure (2.4), (2.5), one has to forestall exposition (5.6), (5.7) by conditions of the non-zero complete CDs of vector functions $\breve{V}^m$ and $\breve{U}_m$:

$$\nabla_k \breve{V}^m \equiv \partial_k \breve{V}^m + G^m_{nk}\breve{V}^n \neq 0; \qquad \nabla_k \breve{U}_m \equiv \partial_k \breve{U}_m - G^n_{mk}\breve{U}_n \neq 0; \tag{5.8}$$

in the opposite case, one has to acknowledge that, RSF form (5.5) turns zero.

*5.1.4. Relation between the RCT and RSF forms.* It is straightforward to derive a relation between two forms, (5.5) and (2.1), based on structure of $\widehat{\mathbf{G}}_{kl}$ given by formula (5.5) and definition of $\widehat{\mathbf{G}}_k$ in (5.5):

$$\widehat{\mathbf{G}}_{kl} \equiv \mathbf{R}_{kl} + \mathbf{T}_{kl}, \tag{5.9}$$

where $\mathbf{T}_{kl}$ appears given by the following formulas:

$$\mathbf{T}_{kl} \equiv T^m_{n;kl} \equiv \hat{\partial}_k \mathbf{T}_l - \hat{\partial}_l \mathbf{T}_k + [\mathbf{T}_k, \mathbf{T}_l], \qquad \mathbf{T}_k \equiv T^m_{nk} \tag{5.10}$$

or:

$$\mathbf{T}_{kl} \equiv \partial_k \mathbf{T}_l - \partial_l \mathbf{T}_k + [\mathbf{\Gamma}_k, \mathbf{T}_l] - [\mathbf{\Gamma}_l, \mathbf{T}_k] + [\mathbf{T}_k, \mathbf{T}_l]. \tag{5.11}$$

Object $\mathbf{T}_{kl}$ can be characterized as a curl-covariant derivative of the tensor-moderator $\mathbf{T}_k$. The latter is presupposed to be a tensor; this certification requires recognition of form $\widehat{\mathbf{G}}_{kl}$ as tensor. Certification of $\widehat{\mathbf{G}}_{kl}$ as tensor follows from consideration of the second complete CD of vector functions.

*5.2. Mediator theorem 2*

*5.2.1. Nullification of form $\widehat{\mathbf{G}}_{kl}$ at existence of mediators* (4.5). By the way, going further, one can easily discover the next, *mediator theorem 2,* which establishes a connection of tensor-moderator $\mathrm{T}_{nk}^m$ to RCT (2.1). Let us consider the alternate second ordinary derivative of the mediator functions $V^m(x)$, $U_m(x)$, subordinate of equations (4.6). Since:

$$\partial_k \partial_l V^m - \partial_l \partial_k V^m \equiv 0,$$

it follows immediately from vector equations (4.6), that:

$$\partial_k(\widehat{\mathrm{G}}_{nl}^m V^n) - \partial_l(\widehat{\mathrm{G}}_{nk}^m V^n) = 0; \tag{5.12}$$

so we receive the following relations:

$$(\partial_k \widehat{\mathrm{G}}_{nl}^m) V^n - (\partial_l \widehat{\mathrm{G}}_{nk}^m) V^n + \widehat{\mathrm{G}}_{nl}^m \partial_k V^n - \widehat{\mathrm{G}}_{nk}^m \partial_l V^n = 0. \tag{5.13}$$

Note that, in distinction from the above shown similar procedure with Christoffel symbols $\Gamma_{nk}^m$, the complete connectedness $\widehat{\mathrm{G}}_{nk}^m$ is not symmetric on down indices. Referring, again, to equations (4.6), we replace derivatives of $V^n$ in (5.13) according to (4.6), and obtain the following relations:

$$(\partial_k \widehat{\mathrm{G}}_{nl}^m - \partial_l \widehat{\mathrm{G}}_{nk}^m + \widehat{\mathrm{G}}_{pk}^m \widehat{\mathrm{G}}_{nl}^p - \widehat{\mathrm{G}}_{pl}^m \widehat{\mathrm{G}}_{nk}^p) V^n = 0. \tag{5.14}$$

Resulting form in brackets before vector $V^n$, the *Riemann supertensor form*, we denoted above $\widehat{\mathrm{G}}_{n;kl}^m$ or $\widehat{\mathbf{G}}_{kl}$, so we can write:

$$\widehat{\mathrm{G}}_{n;kl}^m V^n = 0. \tag{5.15}$$

The correspondent procedure applied to equation for covariant mediator $U_m$:

$$\widehat{\nabla}_k U_m \equiv \partial_k U_m - \widehat{\mathrm{G}}_{mk}^n U_n = 0, \tag{5.16}$$

leads to a similar result:

$$\widehat{\mathrm{G}}_{m;kl}^n U_n = 0. \tag{5.17}$$

Equations (5.15) and (5.17) can be referred to arbitrary direction of the autonomic vector objects $V^m$ and $U_m$ at arbitrary point and overall components of form $\widehat{\mathbf{G}}_{kl}$, so we have to conclude:

$$\widehat{\mathbf{G}}_{kl} \equiv \mathbf{R}_{kl} + \mathbf{T}_{kl} = 0. \tag{5.18}$$

We call phenomenon of nullification of $\widehat{\mathbf{G}}_{kl}$ the *RSF annihilation*.

*5.2.2. Nullification of form $\widehat{\mathbf{G}}_{kl}$ as annihilation of the Riemann supertensor form.* Similar to the case of exposition of RCT from procedure of the alternate second PCD (2.4) – (2.5), we note that, issue of certification of the RSF as geometrical object is conjugate with the following two constraints.

1) Exposition of the RSF form (5.9) as a non-zero tensor from procedure (5.6), (5.7) requires existence of vector functions with a non-zero *complete covariant derivative (CCD)*.

2) On the other hand, regardless to possible existence of the such ones, existence of RSF as a non-zero geometrical object (i.e. tensor) of the Riemannian geometry is not compatible with introduction of autonomic vector functions of zero CCD (4.6), – *mediators*, dedicated to serve exposition of the Riemann-Christoffel tensor (2.1).

We call phenomenon of nullification of the $\widehat{\mathbf{G}}_{kl}$ form *annihilation of the RSF*.

*5.3. RCT as driver for the tensor-moderator*

*5.3.1. Connection of moderator to RCT.* At a first glance, with result (5.18) we received nothing new besides the second zero. However, there actually occurs a crucially constructive internal content in the nullification of the RSF (5.18). Namely, using representation (5.9) for the RSF, we find from (5.18) that, the incorporated tensor $\mathbf{T}_k$ can be viewed driven by the RCT (2.1):

$$\widehat{\mathbf{G}}_{kl} = \mathbf{R}_{kl} + \mathbf{T}_{kl} \Longrightarrow 0; \quad \rightarrow \quad \mathbf{T}_{kl} = -\mathbf{R}_{kl}, \tag{5.19}$$

i.e., according to disclosure (5.10), (5.11):

$$\hat{\partial}_k \mathbf{T}_l - \hat{\partial}_l \mathbf{T}_k + [\mathbf{T}_k, \mathbf{T}_l] = -\mathbf{R}_{kl} \tag{5.20}$$

or:

$$\partial_k \mathbf{T}_l - \partial_l \mathbf{T}_k + [\mathbf{\Gamma}_k, \mathbf{T}_l] - [\mathbf{\Gamma}_l, \mathbf{T}_k] + [\mathbf{T}_k, \mathbf{T}_l] = -\mathbf{R}_{kl}. \tag{5.21}$$

*5.3.2. Resume on the existence and actuality of the tensor-moderator.* We thus have reached a realization of that, incorporating asymmetric tensor-moderator $\mathrm{T}^m_{nk}$ in the connectedness as addition to the Christoffel symbols appears not only possible but necessary extension of the connectedness for utilization of Riemannian geometry as a consistent geometrical system beyond the Euclidian one. Investigation of equation (5.21) towards setting a solution $\mathbf{T}_k$ as functional of $\mathbf{R}_{kl}$ etc. goes beyond the scope of this paper. What can be outlined immediately, is that, there is no indication of that, either one of the two symmetry party of tensor $\mathrm{T}^m_{nk}$, *torsion* and *gravitensor*, might be cancelled to zero. Thinking so far formally and in general, one may consider that, asymmetry of the tensor-moderator $\mathrm{T}^m_{nk}$ on covariant indexes $n$ and $k$ is due to the a background distinction in role of these two indices in equations (5.21) connecting TM to the Riemann-Christoffel tensor. Role and contributions of both party of the TM in the gravitation and geodesics will be elicited below in principal aspects.

### § 6. Dynamics and invariants of vectors-mediators

Here we will point geometrical and dynamical properties of the above introduced vector-mediators.

*6.1. Characterization of the mediators*

The mediators are the contravariant and covariant vector functions $V^m(x)$ and $U_m(x)$, respectively, subordinates of covariant autonomic equations in first derivatives (4.6). An individual mediator, solution of these equations, is determined by its direction at a point. In $N$ dimensions manifold one may choose $N$ independent particular directions at a point as basis; the rest (number of which is infinite) arrive the linear superpositions of the chosen basic ones.

*6.2. Mediator's invariants*

Scalar product $V^m U_m$ arrives an absolute invariant, namely:

1) it is invariant of transformations of the coordinates $x$:

$$V^m U_m = inv ; \qquad (6.1)$$

2) it is also an absolute invariant i.e. constant over the space, as this follows immediately from equations (4.6):

$$\partial_k (V^m U_m) = 0; \quad \rightarrow \quad V^m U_m = const. \qquad (6.2)$$

*6.3. Reduction of the mediator's affine duality to the metrical one*

Affine duality of a mediator couple can be reduced effectively to a metrical one by introduction of the moderate metric tensor $\widehat{w}_{mn}$:

$$U_m \Rightarrow \widehat{w}_{mn} V^n \equiv \widehat{V}_m . \qquad (6.3)$$

Then, moderate metric must satisfy equations with moderate connectedness $\widehat{G}^m_{nk}$ (4.6):

$$\widehat{\nabla}_k \widehat{w}_{mn} \equiv \partial_k \widehat{w}_{mn} - \widehat{G}^l_{mk} \widehat{w}_{ln} - \widehat{G}^l_{nk} \widehat{w}_{lm} = 0 \qquad (6.4)$$

*6.4. Renormalization of metric in dynamics*

Let us denote $\lambda_{mn}$ shift in metric tensor due to the tensor-moderator:

$$\widehat{w}_{mn} = w_{mn} + \lambda_{mn}; \qquad (6.5)$$

then, taking into account that, $\hat{\partial}_k w_{mn} = 0$, we obtain the following equation for $\lambda_{mn}$:

$$\widehat{\nabla}_k \lambda_{mn} = T^l_{mk} w_{ln} + T^l_{nk} w_{lm} . \qquad (6.6)$$

We call renormalized metric $\widehat{w}_{mn}$ the *dynamical* one.

*6.5. Relative status of the contra- and co-variant mediators*

The contravariant mediators can be associated with tangent vectors of special world lines $x^m(\tau)$:

$$\frac{d}{d\tau} x^m(\tau) = V^m(x), \qquad (6.7)$$

while the covariant one, $\widehat{V}_m$ , can be regard only as super*metrical image* of $V^m$:

$$\widehat{V}_m \equiv \widehat{w}_{mn} V^n. \qquad (6.8)$$

*6.6. Scalar product of mediators as an absolute invariant*

We can now define invariant scalar product of the contravariant mediators:

$$(V_a V_b) \equiv \widehat{w}_{mn} V^m_a V^n_b; \qquad \partial_k (V_a V_b) = \partial_k (V^m_a \widehat{V}_{bm}) = 0; \quad \rightarrow \quad (V_a V_b) = const \qquad (6.9)$$

*6.7. Mediator invariant norm.* With renormalized metric tensor, the constant *mediator norm* can be introduced, as well:

$$\mathbf{V}^2 \equiv \widehat{w}_{mn} V^m V^n = inv = const \qquad (6.10)$$

*6.8. Orthogonal mediators*

We call two mediators orthogonal, once their scalar product is equal zero:

$$\widehat{w}_{mn} V_a^m V_b^n = 0. \tag{6.11}$$

Remind that, in view of scalar product constancy over space, it will be zero everywhere over the space, once it has been set to zero at a point.

*6.9. Mediator's orthogonal frame.* The above mentioned collection of basic mediators can be an ensemble of $N$ orthogonal mediators $V_i^m$:

$$\widehat{\nabla}_k V_i^n = 0; \qquad \widehat{w}_{mn} V_i^m V_j^n = 0, \; j \neq i \; ; \qquad \boldsymbol{V}_i^2 \equiv \widehat{w}_{mn} V_i^m V_i^n = const \tag{6.12}$$

*6.10. Mediator expansion*

An arbitrary mediator can be represented as constant superposition of $N$ orthogonal mediators:

$$V^m = C^i V_i^m \; ; \qquad C^i = \frac{\widehat{w}_{mn} V^m V_i^n}{\boldsymbol{V}_i^2} = const. \tag{6.13}$$

*6.11. Issue of the VM geometrical meaning*

Above we introduced a category of vector-mediators (VM) as an element of a covariant differential geometry inquired to substantiate the Riemann-Christoffel form (2.1) (RCT) as a non-zero tensor according to procedure of exposition (). [*)] Setting the VMs as autonomic vector functions with a non-zero *primal covariant derivatives* (PCD) $\hat{\partial}_k V^m, \hat{\partial}_k U_m$:

$$\hat{\partial}_k V^m \equiv \partial_k V^m + \Gamma_{nk}^m V^n = - \mathrm{T}_{nk}^m V^n; \qquad \hat{\partial}_k U_m \equiv \partial_k U_m - \Gamma_{mk}^n U_n = \mathrm{T}_{mk}^n U_n \, , \tag{6.14}$$

we derived law (5.20) of tensor-moderator (TM) $\mathrm{T}_{nk}^m$ connection to RCT. Form of differential law of the VMs, $V^m$, $U_m$ thus distinguishes from that of vectors $g^m$, $g_m$ of ES (3.12) by term with tensor-moderator TM $\mathrm{T}_{mk}^n$ generated by the RCT. So contravariant mediator $V^m$ of the SRS can be regard as analog of vectors of straights $g^m$ of ES. This observation leads to a presumption that, vector-mediators $V^m$ being certified as autonomic vector functions of a non-zero PCD (6.14), actually might be recognized associated with geodesic lines of SRS, − being the *tangent vectors* of the latter.

## IV. Geodesics in Riemannian geometry

### § 7. Correspondence principles of a geodesic concept

*7.1. Preamble.* The *geodesics* i.e. *geodesic lines* (GLs) have been generally predestined in Riemannian geometry to play a role similar to that of the *straight lines* or, simply, *straights* in Euclidian (*flat geometry*) space. As it has been illustrated above, in Euclidian space (ES) vector $g^m$, direction of a straight, arrives a function of $x$ in terms of a curved coordinate frame. Now, considering picture in a strict Riemannian space (SRS), it should be emphasized that, not just the covariance requirement itself but, first of all, the geometry properties of SRS unavoidably obey one to imply use of the curved coordinate frames as fundamentally inherent to the SRS (i.e. *Riemannian*) geometry. It should be noted in this connection that, a correspondent background distinction of the SRS from ES consists in a circumstance that, metric tensor dependence of the coordinates cannot be factorized similar to certification of the Euclidian geometry (2.11). Therefore, in the SRS geometry with its curvature tensor (2.1), this "distortion"(represented by the non-zero RCT) cannot be removed by choice of an "appropriate" coordinate frame. Thus, category of the *tangent vector* of a geodesic line appears forcefully yet naturally promoted to the category of a *vector function* $G^m(x)$, subordinate of a certain differential law (*geodesic vector law*, GVL). Once the GVL has been derived and stated, equation for geodesic lines follows it immediately:

$$x^m \rightarrow x^m(\tau); \quad \frac{dx^m}{d\tau} = G^m(x); \quad \frac{d^2x^m}{d\tau^2} \equiv \frac{d}{d\tau}G^m(x) \equiv \frac{dx^k}{d\tau}\partial_k G^m \equiv G^k \partial_k G^m; \quad (7.1)$$

here, $\tau$ is a *canonical parameter* []. Introduction of the latter appears a convenient general manner of representing the world lines along which the coordinates are connected by a specific law.

One may generally presume that, condition of the *correspondence* to the Euclidian geometry (EG) in corporation with requirements of the covariance and consistence of the GVL with respect of existence of RCT should navigate the deriving of equation for *geodesic vector function* (GVF) in Riemannian geometry and GTR.

Requirements of the correspondence and consistence to the formulating the geodesics, in our view, should include the following three aspects.

*7.2. General stature of the GVL*

Geodesic vector law for tangent vectors of lines in a strict Riemannian space should be formulated as an autonomic differential equation for vector functions of coordinates, a further going covariant extension of equations for straights in a curved frame of an Euclidian space.

*7.3. Compatibility with the RCT*

Differential law for tangent vectors of the geodesics must be compatible with existence of the Riemann-Christoffel curvature tensor (2.1).

*7.4. Asymptotic authenticity of the geodesics to straights of Euclidian space*

Distinction of the tangent vector law from equation for directions of straights in ES (37) should vanish together with the RCT.

Concerning the latter claim, we have to stress that, in the limit $\mathbf{R}_{kl} \Longrightarrow 0$, form of the geodesic law may tend to equations (3.12), (3.21) but not obligatorily directly to the rudimentary equations (3.1), (3.2) for straights in Euclidian space, − according to the exposed above in paragraph **3**.

## 8. Profiling the geodesic vector law

*8.1. GVL as equation for an autonomic contravariant vector object*

As mentioned above, GVL should be equation for an autonomic contravariant vector object, since orientation (i.e. direction) of GVF is not predetermined by an external factor. This implies the following presupposed form of equations for geodesic vector function $G^m(x)$:

$$\partial_k G^m = -Z^m_{nk}(x)G^n, \quad or: \quad \partial_k G^m + Z^m_{nk}G^n = 0; \quad (8.1)$$

here $Z^m_{nk}$ are $N$ matrices on indices $m, n$, functions of coordinates.

*8.2. Covariance of the GVL form*

Since the GVL must be covariant, one may presuppose the following general disclosure of matrices $Z^m_{nk}$:

$$Z_{nk}^m \Rightarrow \Gamma_{nk}^m + Y_{nk}^m, \qquad (8.2)$$

where $\Gamma_{nk}^m$ are Christoffel symbols (1.24), and $Y_{nk}^m$ is an asymmetric tensor. The latter can be inquired and specified in view of other necessary features of the geometrical system.

*8.3. Test of the primal utilization of the geodesics*

Above in paragraphs *3.1.* , *3.2.* we demonstrated representation of straights of Euclidian space in terms of a curved frame of coordinates $x^k$:

$$\frac{dx^m}{d\tau} = g^m(x); \quad \frac{dg^m}{d\tau} = g^k \partial_k g^m; \quad \partial_k g^m + \bar{\Gamma}_{nk}^m g^n = 0, \quad \rightarrow \quad \frac{d^2 x^m}{d\tau^2} = -\bar{\Gamma}_{nk}^m \frac{dx^n}{d\tau}\frac{dx^k}{d\tau}, \qquad (8.3)$$

where $\bar{\Gamma}_{nk}^m$ is Christoffel symbols (2.12) at metric tensor of ES in curved frame $\bar{w}_{lm}(x)$ given by formula (2.11). Remind that, curvature tensor (2.1) remains zero there. Turning to the case of Riemannian geometry, one may examine a proposition that, geodesic law could be formulated here by similar equations, but with replacement $\bar{w}_{lm} \rightarrow w_{lm}$ , at no presupposed specification of metric tensor $w_{lm}$ which, presumably, is associated now with the RCT (2.1). However, we have observed in paragraph *3.2.* that, the law:

$$\partial_k g^m + \Gamma_{nk}^m g^n = 0 \qquad (8.4)$$

is not compatible with existence of a non-zero RCT form (2.1) i.e. with Riemannian geometry. In fact, law (8.4) still belong realm of the of the Euclidian geometry, resulting in solution (2.11) for metric tensor:

$$w_{lm}(x) \Rightarrow \bar{w}_{lm}(x), \qquad (8.5)$$

as it has been shown in paragraph *3.2.* .

*8.4. Tensor $Y_{nk}^m$ as identity of the tensor-moderator $T_{nk}^m$*

Presence of tensor party, $Y_{nk}^m$ , of object $Z_{nk}^m$ can be motivated by consideration of a consistence of ensemble of the objects those are constituting the establishment of the Riemannian geometry (RG) system. Certainly, the geodesic vector function $G^m(x)$ should belong the RG establishment being an irreducible object i.e. object not structured on other objects. However, its properties should be compatible with existence and necessary properties of other independent objects. First of all, presence of tensor $Y_{nk}^m$ in the connection matrices $Z_{nk}^m$, obviously, is required in order to have a possibility to align existence of the GVFs with existence of the RCT. Strictly speaking, tensor $Y_{nk}^m$ must be introduced in equation for GVF, since RG excludes existence of vector functions with zero PCD as shown in paragraph *3.2.* . Furthermore, this tensor is naturally and uniquely recognizable as identity of tensor-moderator $T_{nk}^m$ of the complete connectedness $\hat{G}_{nk}^m$ . Hence, we do not need to search for a new triadic object $Z_{nk}^m$, other than $\hat{G}_{nk}^m$.

*8.5. GVFs equations*

We thus have figured out that, logic of the meaning and genesis of the *geodesic vector function* strained with principle of correspondence to directions of the straights in Euclidian space leads to equation for GVF of a simple formula: the *complete covariant derivative* (CCD) of GVF taken in its generally resumed definition (8.1), (8.2 ) should be equal zero, namely:

$$\partial_k G^m + \hat{G}_{nk}^m G^n = 0, \quad or: \quad \partial_k G^m + \Gamma_{nk}^m G^n = -T_{nk}^m G^n. \qquad (8.6)$$

*8.6. Contravariant mediators as geodesic vector functions of Riemannian geometry*

Above at introduction of the RCT we focused on a requirement that, there should exist a dual pair of vector functions, PCD of which should not be equal zero (see (*2.6*)). We then can identify the above introduced *contravariant* mediators $V^m(x)$ with geodesic vector functions $G^m(x)$ of the Riemannian geometry. Vice versa, GVFs (8.6) plays role of the autonomic mediators for introduction of the RCT through the exposition (2.4), (2.5).

*8.7. Geodesic lines of Riemannian geometry*

*8.7.1. Equation of the geodesic lines.* Once equation (8.6) for GVs has been formulated, transition to equation for *geodesic lines* (GLs) is elementary straightforward according to the above shown exposition (7.1). Referring then to equation (8.6), we obtain:

$$G^k \partial_k G^m = -\widehat{G}^m_{nk}(x) G^n G^k = -\overline{\overline{G}}^m_{nk} G^n G^k; \qquad \overline{\overline{G}}^m_{nk} \equiv \Gamma^m_{nk} + \overline{\overline{T}}^m_{nk}; \tag{8.7}$$

here $\overline{\overline{T}}^m_{nk}$ is the even-symmetric part of the tensor-moderator $T^m_{nk}$, we call the *gravitensor*. The skew-symmetric part, $\overline{T}^m_{nk}$, *torsion*, does not contribute in summation over indices $n, k$ in (8.7). In result, we obtain the following equations of the geodesic lines:

$$\frac{dG^m}{d\tau} = -\widehat{G}^m_{nk}(x) G^n G^k \quad \rightarrow \quad \frac{d^2 x^m}{d\tau^2} + (\Gamma^m_{nk} + \overline{\overline{T}}^m_{nk}) \frac{dx^n}{d\tau} \frac{dx^k}{d\tau} = 0. \tag{8.8}$$

Note, however, that, generation of the gravitensor $\overline{\overline{T}}^m_{nk}$ by the RCT is influenced by the torsion which borns together with the gravitensor in one process generation of tensor-moderator $T^m_{nk}$ according to equations (5.20).

*8.7.2. Local equivalence of the Riemannian and Euclidian geometries.* This property of the Riemannian geometry certified with geodesics (8.8), manifests with a possibility of a local nullification of the connectedness $\overline{\overline{G}}^m_{nk}$ by a certain transformation of the coordinates. Namely, referring to transformation formula (1.21) with requirement to a particular transformation of the coordinates:

$$\overline{\overline{G}}^m_{nk}(x) \quad \rightarrow \quad \overline{\overline{G}}^{m'}_{n'k'} = 0, \tag{8.9}$$

we obtain the following condition for transformation matrix $A^{m'}_m$ [3-9]:

$$A^{m'}_m \overline{\overline{G}}^m_{nk} - \partial_k A^{m'}_n = 0, \quad \text{or}: \quad \frac{\partial^2 x^{m'}}{\partial x^n \partial x^k} = (\Gamma^m_{nk} + \overline{\overline{T}}^m_{nk}) \frac{\partial x^{m'}}{\partial x^m}, \tag{8.10}$$

which can be satisfied at a point arbitrary chosen in advance, and along a line [8].

## VI. Gravitation in GTR

### § 9. Gravitation concepts of the GTR: review and notes

*9.1. The Einstein – Hilbert metric concept of the GTR*

Riemannian geometry itself does not treat question about origin (genesis) of deviation of metric of the World from that of the flat Euclid – Minkowski space. Raising and solving this question is the central aspect of general theory of relativity (GTR) created by A. Einstein. Covariant relativistic concept of the gravitation field (GF) introduced by A. Einstein and D. Hilbert as a 4-fold covariant extension of the I. Newton' gravitation law in the Riemann – Minkowski space consists of a statement of the metric tensor $w_{ik}$ coupling to the energy – momentum tensor $M_{ik}$ (our notation) of the *matter* [1-4]:

$$R_{ik} + \frac{1}{2}w_{ik}R = \kappa M_{ik}; \qquad R_{ik} \equiv R^{l}_{i;kl} = R_{ki}; \qquad R \equiv w^{ik}R_{ik}; \tag{9.1}$$

here $R^{m}_{i;kl}$ is the *curvature tensor* (2.1), and $\kappa$ is the *gravitation constant*. This equation substantiates metric tensor $w_{ik}$ as function of the coordinates, connecting it to the *matter*.

*9.2. Conventional formulations of the geodesic law in GTR*

*9.2.1. Concept of the geodesics as extension of the Galilean law of the inertial motion.* The geodesics i.e. geodesic lines (GLs) have been generally predestinated in the Riemannian geometry to play a role similar to that of the *straight lines* or simply *straights* in Euclidian (*flat geometry*) space. When creating the GTR, A. Einstein incepted there a correspondent principle by his genius statement that, particles in a gravitation field move along the geodesic lines of the Riemann-Minkowski geometry. It has to be thought that, his idea was based on consideration of geodesic law in particle dynamics as covariant generalization of the Galileian law of *inertial motion* in space free of the gravitational and other interaction fields; in turn, such a thinking might be regard as directed by his *principle of the equivalence* [1]. The essence of this issue consists, however, of a question, what a specific formulation might promote such a fundamental proposition to a certain mathematical utilization in terms of the Riemannian geometry.

There are two well-known approaches to deriving the law of the geodesic lines in GTR:

1. Primal covariant extension of the Galilean law of inertia.

2. Resorting to variation principle of minimum interval with variable metric.

*9.2.2. Primal covariant extension of the inertia law.* This approach represents a direct exposure of the geodesic law through Christoffel symbols based on the following statements.

1. Effect of the gravitation is enclosed in the metric tensor change with the space-time coordinates.

2. More specific, gravitation field (GF) is represented immediately and in whole by the Christoffel symbols.

Therefore, according to these statements, formulation of the law of particle motion in GF can be enclosed in replacement of the particle 4-velocity differential $du^m$ in the Galilean law of inertia by the *primal* (our word) covariant differential $Du^m$ in terms of a curved coordinate frame [4] (symbol $D$ of covariant differential in our notation is $\hat{\partial}$):

$$du^m = 0 \quad \rightarrow \quad Du^m \equiv du^m + \Gamma^{m}_{nk}u^n dx^k = 0; \tag{9.2}$$

then, using definition:

$$u^m = \frac{dx^m}{d\tau} \tag{9.3}$$

(here and below parameter $\tau$ represents integral of time in the rest frame of a particle []) ones write the relations:

$$\frac{du^m}{d\tau} + \Gamma^{m}_{nk}u^n u^k = 0; \qquad \frac{dx^m}{d\tau} = u^m \tag{9.4}$$

which can be rewritten as differential equations for $x^m(\tau)$:

$$\frac{d^2 x^m}{d\tau^2} + \Gamma^{m}_{nk}(x)\frac{dx^n}{d\tau}\frac{dx^k}{d\tau} = 0. \tag{9.5}$$

*9.2.3. Extreme Action method of deriving the geodesic law.* Initial position of resorting to the *Extreme Action principle* as a method to derive covariant law of particle motion, in case of the gravitation field consists of a statement that, one may use the same action integral form as for particle motion in absence of the gravitation field, based on a general proposition (regard as an underlying genetic discovery) that, effect of the gravitation consists of changing of the space-time metric with the coordinates []. Variation procedure [] (see Appendix A3) results in the conventional geodesic equations (9.5).

*9.3. Notes to the conventional formulations of geodesics in GTR*

*9.3.1. Common notes.*

1. First, one may observe that, equation (9.5) coincides with equation for straights in a curved frame of Euclidian space (3.21). This observation itself does not arrive a proving argument, but it urges to undertake a comprehensive disclosing analysis of this equation.

Let us write equations (9.5) in the foregoing form as equations for $x^m$ and $u^m$:

$$\frac{du^m}{d\tau} + \Gamma^m_{nk} u^n u^k = 0; \qquad \frac{dx^m}{d\tau} = u^m \tag{9.6}$$

In a curved coordinate frame inherent to GTR, particle velocity as a direction becomes function of coordinates (time is one of those), then we can write:

$$\frac{d}{d\tau} u^m(x) \equiv u^k \partial_k u^m; \tag{9.7}$$

finally, we receive equations for $u^m(x)$ in a curved frame, with summation on index $k$:

$$u^k (\partial_k u^m + \Gamma^m_{nk} u^n) \equiv u^k \hat{\partial}_k u^m = 0.$$

Further, one may consider two options.

A). $\qquad \hat{\partial}_k u^m \equiv \partial_k u^m + \Gamma^m_{nk} u^n = 0 ;$ \qquad (9.8)

this will lead us to the outcome: RCT= 0, proved above in paragraph *3.2.* .

B). $\qquad \hat{\partial}_k u^m \neq 0,$ \qquad (9.9)

but, nevertheless,

$$u^k \hat{\partial}_k u^m = 0. \tag{9.10}$$

With this assumption, one should take into account two circumstances:

1. Since covariant derivative $\hat{\partial}_k u^m \neq 0$ is tensor, it then should be equalized with some tensor $X^m_k$.

2. Tensor $X^m_k$ should be a conversion of some triadic tensor $X^m_{nk}$ with $u^n$, in view of the autonomic status of particle 4-velocity $u^k$.

Assumption B) thus leads to a form of equation with complete covariant derivative (8.8):

$$\hat{\partial}_k u^m = -X^m_{nk} u^n , \tag{9.11}$$

but under the following condition:

$$\overline{\overline{X}}^m_{nk} u^n u^k = 0; \tag{9.12}$$

here $\overline{\overline{X}}^m_{nk}$ is part of tensor $X^m_{nk}$ even-symmetric on the down indices. By the way, as an elementary analysis conducted above in paragraph *3.2.* had shown, equation (9.11) leads to connection (5.20) of both parts of tensor $X^m_{nk}$ to RCT, at no correlation with the geodesic vector (i.e. particle 4-velocity $u^k$). Thus, there is no way to satisfy condition (9.12) at a non-zero $\overline{\overline{X}}^m_{nk}$. We thus return back to option (9.8) incompatible with Riemannian geometry.

*9.3.2. Notes specifically to the primal concept of the geodesics* (9.2).

1. Replacement (9.2) is conventionally regard as an immediate transition to the Riemannian geometry (i.e. considered as effect of the gravitation to metric []). Meanwhile, metric tensor becomes variable in space, in particular, at arbitrary non-degenerate, non-linear transformation of coordinates from Cartesian frame in Euclidian space; the associated Christoffel symbols (CSs) arrive there, but RCT still be zero over the space (see section …). So, variability of metric tensor in space (or presence of the CSs) itself cannot be unconditionally considered a signature of the Riemannian geometry (i.e. presence of RCT) of the space.

2. On the other hand, presence of CSs in differential relatitions does designate, with inevitability, factual reference to equations (9.8) with explicit partial derivatives of a vector function of coordinates. Remind and underline that, CSs $\Gamma^m_{nk}$ as a mixed valence 3 geometrical object with their transformation law (1.21) arrive when ones consider transformation of partial derivatives of a vector function (at curved transformations of coordinates) under a requirement that, partial *covariant* derivatives of vectors:

$$\hat{\partial}_k V^m \equiv \partial_k V^m + \Gamma^m_{nk} V^n \tag{9.13}$$

should transform as tensor. In turn, we stress that, particle 4-velocity in terms of a curved frame appears generally a function of 4- coordinates regardless of the type of the space geometry. Realization of this circumstance unavoidably returns one to a normal representation of the ordinary differential of particle 4-velocity along a line $x^k(\tau)$, i.e. one can write:

$$du^m(x) \equiv \frac{\partial u^m(x)}{\partial x^k} dx^k. \tag{9.14}$$

Taking then differentials $dx^k$ along a world line $x^k(\tau)$, we receive:

$$du^m = \frac{\partial u^m}{\partial x^k} \frac{dx^k}{d\tau} d\tau \equiv (\partial_k u^m) u^k d\tau; \quad \rightarrow \quad \frac{du^m}{d\tau} = u^k \partial_k u^m. \tag{9.15}$$

Writing covariant differential along a line in relation (9.2) actually implies summation of the projected partial derivatives by formula (9.15). Short manner of (9.2) shadows disclosure (9.15) but does not grant one with a reason to ignore differential equations (9.8) which, in fact, stands as a genetic code of the suggested geodesic line equation (9.5). Again, we underline that, the pointed backward disclosure of the differential $du^m$ in (9.2) appears completely legitimate, consistent, true and necessary in view of the reason that, vector (i.e. direction) of particle velocity considered in terms of a curved coordinate frame of the Riemannian geometry acquires gradients in a vicinity of particle position; therefore, it can be treated as vector function of the coordinates $x^k$, and actually arrives so.

3. As a general conclusive characterization of the content and sense of the covariant differential $Du^m$, it represents only a necessary first step in formulation of the geodesic law in Riemannian geometry, namely, formulation of a covariant derivative of tangent vectors of the geodesic lines in a curved coordinate frame.

However, equating it to zero cannot deliver a new physical content beyond the inertia law in Euclidian space. Response to the curvature impact inherent to Riemannian geometry of the gravitation becomes acquired only when covariant differential operator $\hat{\partial}_k$ gets confronting with tensor-moderator $T_{nk}^m$ generated by the RCT, as it has been outlined from the above presented analysis.

*9.3.3. Notes to the action integral method.* When considering question about adequacy of resorting to the EAP, We note that, the known formulation [ ] does not draw constraints of Riemannian geometry associated with existence of the Riemann-Christoffel tensor. Therefore, the derived equation (9.5) coincides with equation for straights in Euclidian space in terms of a curved frame (3.9) or (3.21). Remind that, null of RCT appears a direct consequence of equation (9.8). By the way, such an outcome looks likely to signify a statement that, the extremum (minimum) of action integral is attained in a flat (Euclid – Minkowski) space, in comparison with a strict Riemannian space of a fixed RCT. Search for a related consistent action integral looks likely to employ the Lagrange method of uncertain multipliers reflecting existence of a non-zero RCT form. Such an inclination goes beyond the scope of this paper. Our consideration of issue of the geodesics in a strict Riemannian space consist of the more immediate arguments and methods referring, in one side, to conceptual parallelism with straights of a Euclidian space in a curved frame, and, in the other side, to existence and genetic origin of the RCT.

*9.3.4. Conclusion of review of the conventional GTR geodesic concept.* We thus are enforced to conclude that, equation (9.5) conventionally assumed in GTR as representing the law of particle motion in gravitation field (the latter is identified with Christoffel symbols), actually does not fit such a designation. In fact, it only expresses the law of inertial motion of particles along straight lines in a flat (i.e. Minkowski's) space-time manifold in terms of a curved frame. In view of this circumstance, substitution of Christoffel symbols exposed from solution of the metric equation (9.1) into formula (8.5) does not attain the attitude of a consistent approach to the problem of description of particle motion in gravitation field. This disposition, certainly, is a reflection of a circumstance that, variation of metric with coordinates is a necessary but not a sufficient condition of transition to the Riemannian geometry. Existence of the Riemann-Christoffel curvature tensor is the only true watershed between the flat and curved geometry spaces and related treats of geometry and gravitational dynamics; so the RCT shall be manifesting in the geodesic law of the GTR.

## VII. The update dynamical concept of the GTR

### § 10. Update theory of the gravitation field and particle dynamics

*10.1. General statement of the law of particle motion in gravitation field*

*10.1.1. Principal thesis (by A. Einstein).* In absence of other forces, particles move along the geodesics of Riemann space of the gravitation field. Particle 4-velocity arrives tangent vector of the geodesic world line.

$$u^m \Rightarrow G^m \equiv \frac{dx^m}{d\tau} \tag{10.1}$$

*10.1.2. Particle 4-velocity in a curved frame of Riemann space of the gravitation field.* 4-vector of the velocity generally appears function of 4 coordinates subordinate of the *geodesic vector law* (GVL) (outline of the present paper):

$$\partial_k u^m + \Gamma_{nk}^m u^n = -T_{nk}^m u^n, \tag{10.2}$$

where $\Gamma^m_{nk} \equiv \mathbf{\Gamma}_k$ are Christoffel symbols (1.24), and $T^m_{nk} \equiv \mathbf{T}_k$ is *tensor-moderator* (1.26) induced by the Riemann-Christoffel curvature tensor $\mathbf{R}_{kl} \equiv R^m_{n;kl}$ (2.1):

$$\partial_k \mathbf{T}_l - \partial_l \mathbf{T}_k + [\mathbf{\Gamma}_k, \mathbf{T}_l] - [\mathbf{\Gamma}_l, \mathbf{T}_k] + [\mathbf{T}_k, \mathbf{T}_l] = -\mathbf{R}_{kl}. \tag{10.3}$$

*10.2. Gravitation field (or complete connectedness)*

$$GF = -\widehat{G}^m_{nk} \equiv -(\Gamma^m_{nk} + T^m_{nk}); \tag{10.4}$$

$$T^m_{nk} \equiv \mathbf{T}_k = \overline{T}^m_{nk} + \overline{\overline{T}}^m_{nk}; \qquad \overline{\overline{T}}^m_{nk} = \overline{\overline{T}}^m_{kn} - (gravitensor); \qquad \overline{T}^m_{nk} = -\overline{T}^m_{kn} - (torsion) \tag{10.5}$$

*10.3. Particle trajectories in GF*

*10.3.1. Definition of a trajectory:*

$$x^m(\tau); \quad \frac{dx^m}{d\tau} = u^m(x) \tag{10.6}$$

*10.3.2. Particle acceleration in gravitation field*

$$\frac{d}{d\tau} u^m(x) \equiv u^k \partial_k u^m = -\widehat{G}^m_{nk} u^n u^k = -\overline{\overline{G}}^m_{nk} u^n u^k; \qquad \overline{\overline{G}}^m_{nk} = \Gamma^m_{nk} + \overline{\overline{T}}^m_{nk} \tag{10.7}$$

*10.3.3. Equations of the geodesics*

$$\frac{d^2 x^m}{d\tau^2} = -\overline{\overline{G}}^m_{nk}(x) \frac{dx^n}{d\tau} \frac{dx^k}{d\tau} \tag{10.8}$$

*10.4. Equivalence principle*

The equivalence principle by A. Einstein, as known, consists of a possibility to point a system of coordinate such that gravitational acceleration becomes nullified at a point or along a line in the 4-fold space-time. Namely, it follows from the transformation law of $\overline{\overline{G}}^m_{nk}$ (1.21) that, it appears so at []:

$$\frac{\partial^2 x^{m'}}{\partial x^n \partial x^k} = \overline{\overline{G}}^m_{nk} \frac{\partial x^{m'}}{\partial x^m}. \tag{10.9}$$

*10.5. Essentiality of the tensor-moderator*

Formulation of a consistent GTR dynamic concept while avoiding the inclusion of the tensor-moderator does not seem possible. Moreover, the Riemannian geometry itself receives a certificate of the existence only after that the moderator is introduced as an addition to the Christoffels in the definition of the connectedness object. Being connected to the Riemann-Christoffel curvature tensor in a quite an entanglement manner, tensor-moderator barely could be reduced either to torsion or the gravitensor; one does not get birth without the other one. Note that, though torsion does not contribute directly in the *gravitational acceleration*, it influences the inducing of the gravitensor by the RCT (i.e. by the curvature) in their mutually correlated arise in the strict Riemannian space.

*10.5. 4-velocity norm*

*10.5.1. 4-velocity norm in the Euclid-Minkowski space:*

$$\boldsymbol{u}^2 = w_{mn}u^m u^n = u_m u^m = const \tag{10.10}$$

*10.5.2. Constant norm of the 4-velocity in gravitation field:*

$$\boldsymbol{u}^2 \equiv \widehat{w}_{mn}u^m u^n \Rightarrow const; \tag{10.11}$$

*10.5.3. Renormalization of metric*

$$\widehat{\nabla}_k \widehat{w}_{mn} = 0$$

$$\widehat{w}_{mn} = w_{mn} + \lambda_{mn}; \tag{10.12}$$

then, taking into account that, $\widehat{\partial}_k w_{mn} = 0$, we obtain the following equation for $\lambda_{mn}$:

$$\widehat{\nabla}_k \lambda_{mn} = T^l_{mk} w_{ln} + T^l_{nk} w_{lm}. \tag{10.13}$$

# VIII. Spin in gravitation field

## § 11. Spin in the update GTR

Consideration of spin dynamics of small objects in external gravitation field presents a significant interest, in particular, in view of effect of torsion; the latter contributes immediately in spin precession due to the gravitation field. Consideration below is limited by drawing basic equations for spin 4-vector; treat in more detail comes beyond the scope of this paper.

*11.1. Spin as 4-vector orthogonal to the 4-velocity*

$$\boldsymbol{S}_{rest} = (\boldsymbol{\sigma}; 0); \quad \boldsymbol{P}_{rest} = (\boldsymbol{0}; mc); \quad \rightarrow \quad (\boldsymbol{PS}) = (\boldsymbol{PS})_{rest} = \boldsymbol{0} \tag{11.1}$$

$$\widehat{w}_{mn} u^m S^n = inv = const = 0 \tag{11.2}$$

*11.2. Spin in a curved frame of GTR as 4-vector-mediator*

$$S^m \Rightarrow S^m(x),$$

*11.2.1. Spin 4-fold precession in GF*

$$\widehat{\nabla}_k S^m = \partial_k S^m + \widehat{G}^m_{nk} S^n = 0 \tag{11.3}$$

*11.2.2. Spin norm in gravitation field*

$$\boldsymbol{S}^2 \equiv \widehat{w}_{mn} S^m S^n; \quad \partial_k \boldsymbol{S}^2 = 0; \quad \boldsymbol{S}^2 = const \tag{11.4}$$

*11.3. Spin in orthogonal geodesic frame*

*11.3.1. Orthogonal co-moving frame.* The above mentioned collection of basic mediators can be an ensemble of 4 orthogonal mediators $V_i^m$, one of which is particle 4-velocity; rest three can be associated with spin degrees of freedom:

$$\widehat{\nabla}_k S_i^n = 0; \qquad \widehat{w}_{mn} u^m S_i^n = 0, \; i = 1,2,3 \,; \qquad \boldsymbol{S}_i^2 \equiv \widehat{w}_{mn} S_i^m S_i^n = const \tag{11.5}$$

*11.3.2. Spin expansion.* Spin 4-vector can be represented as constant superposition of *three* orthogonal 4-vectors transvers to the 4-velocity:

$$S^m(x) = C^i S_i^m(x); \qquad C^i = \frac{\widehat{w}_{mn} S^m S_i^n}{\boldsymbol{S}_i^2} = const. \tag{11.6}$$

*11.4. Spin dynamics along the particle trajectories*

Equations for spin precession along particle trajectory can be obtained by differentiation of spin 4-vector function $S^m(x)$ along particle trajectory (geodesic) $x(\tau)$:

$$S^m(x) \Rightarrow S^m[x(\tau)]; \tag{11.7}$$

$$\frac{dS^m}{d\tau} \equiv \frac{\partial S^m}{\partial x^k} \frac{dx^k}{d\tau} = -\widehat{G}_{nk}^m(x) S^n u^k(x); \qquad x \Rightarrow x(\tau); \tag{11.8}$$

finally, these equations can be written in a canonical form as follows:

$$\frac{dS^m}{d\tau} = -G_n^m(\tau) S^n; \qquad G_n^m(\tau) \equiv \widehat{G}_{nk}^m(x) u^k(x); \qquad x \Rightarrow x(\tau). \tag{11.9}$$

Note that, both invariants (11.5) are preserved in this dynamics:

$$\frac{d}{d\tau}(u^m S_m) = u^k \partial_k (u^m S_m) = 0 \,; \qquad \frac{d}{d\tau}(S^m S_m) = u^k \partial_k (S^m S_m) = 0. \tag{11.10}$$

# IX. Resume

## § 12. Summary and Conclusions

*12.1. Summary*

*12.1.1.* The elementary differential analysis of the basic establishment of GTR as theory of the gravitation conducted in the present paper has led to a conclusion that, the conventional certification of the geodesics in GTR is incompatible with the existence of the Riemann-Christoffel curvature tensor.

*12.1.2.* Proposed resolution of the pointed contradiction consists of a natural extension of the Christoffel symbols in the dynamical definition of the geodesic vector function to a complete form of the connectedness incorporating asymmetric triadic tensor as a *moderator* term.

*12.1.3.* Similar analysis of the geodesic law, managed with the enriched connectedness, allows one to derive equation connecting the tensor-moderator to the Riemann-Christoffel tensor, in this way closing the pointed problem of an inconsistence in the dynamic concept of the GTR.

*12.1.4.* Tensor-moderator with its two parts, *torsion* and *gravitensor* (the even-symmetric part, contributing in the gravitation force) appears an indispensable integral structural element of the upgrade dynamical concept of the GTR.

*12.1.5.* We underline that, occurrence of the *gravitensor* as a symmetric tensor addition to the Christoffels in the gravitational force does not abolish the *equivalence principle* posed by A. Einstein in foundation of the General Theory of Relativity.

*12.1.6.* Change in the gravitational dynamic law of GTR considered in this paper does not touch the form of the Einstein-Hilbert equation that connects metric tensor to the energy-momentum tensor of the matter.

*12.1.7.* When considering the phenomenon of torsion in the gravitation theory, we have to distinct between the *gravitation field* and gravitation force: the first one shows a direct torsion term, the second does not. However, torsion does effect birth of the gravitensor (and vice versa) in the process of inducing both of them by the curvature tensor.

*12.1.8.* In this paper we also derived equations for particle spin (either elementary or macro) as for a 4-vector orthogonal to the particle 4-momentum. These equations contain terms with tensor-moderator in its complete form including torsion.

*12.1.9.* Tensor-moderator results also in renormalization of metric tensor when determining the invariant relativistic squares of the 4-vectors of particle momentum and spin in gravitation field.

*12.1.10.* The elicited contradiction in formulation of the geodesic concept concerns not only the gravitation dynamic law of the GTR, but equally the Riemannian geodesic conception itself as a mathematical base in the treats and applications when ones would restrict the certification of the *connectedness* object by the Christoffel symbols. Namely, we state that, a consistent certification of the geodesics can be realized only with implication of the tensor-moderator $T_{nk}^m$ in the dynamics of the geodesic vectors. Tensor-moderator with its two party, torsion and gravitensor, arrives an integral element of the Riemannian geometry certified with the existence of the Riemann-Christoffel *curvature tensor* (RCT). A uniqueness of this disposition consists of such an essential circumstance that, tensor-moderator appears generated by the RCT.

*12.2. Conclusions*

The pointed out an inconsistence in the dynamical aspects of the conventional gravitation concept of the GTR, and the described natural way to refine the definition of the gravitation field, geodesics and spin dynamics, in our understanding, may serve as a sign of that, the gravitation concept of GTR in the existing conventional formulation is essentially incomplete, thus suggesting further explorations of General Theory of Relativity and gravitation in principles and applications.

Equation for tensor-moderator connecting this tensor to the Riemann-Christoffel tensor is the main constructive result of the presented analysis and considered update of the gravitational dynamics of the GTR. In particular, and, perhaps, first of all, the elicited presence of torsion in structure of the gravitational dynamics may lead potentially to investigations of possible effects of torsion' interference with the electrodynamics and models of the strong and weak interactions [10]. Torsion impact to mechanics of spinning objects may be of a significant interest as well.

In this paper we do not touch questions of confrontation of the presented update of the GTR dynamical concept with related observations in astrophysics [1-7, 11]; the correspondent analysis requires special considerations.

*Acknowledgements* It is my great pleasure to thank Anatoliy Kondratenko, Victor Mokeev, Dmitriy Ryutov, Roald Sagdeev, Edward Shuryak, Daniel Usikov and Arkady Vainshtein for useful discussions.

*Authored by Jefferson Science Associates, LLC under U.S. DOE Contract No. DE-AC05-06OR23177. The U.S. Government retains a non-exclusive, paid-up, irrevocable, world-wide license to publish or reproduce this manuscript for U.S. Government purposes.*